\documentclass{article}

\usepackage[right= 0.75in, left= 0.75in, top=1.05in, bottom=1.15in]{geometry}

\usepackage{lineno,hyperref}
\modulolinenumbers[5]

\usepackage{amsmath}
\usepackage{amsfonts}
\usepackage{bm} 
\usepackage{booktabs} 
\usepackage{array} 
\usepackage{paralist} 
\usepackage{verbatim} 
\usepackage{subfig} 
\usepackage[utf8]{inputenc}
\usepackage[affil-it,auth-sc]{authblk}
\usepackage[round,numbers,sort&compress]{natbib}
\usepackage{accents} 
\usepackage{siunitx}
\usepackage{tabularx}
\usepackage{graphicx}
\usepackage{color}

\DeclareMathOperator{\divg}{\mbox{div}} 

\newcommand{\mb}[1]{\mathbf{#1}}

\newcommand{\mc}[1]{\mathcal{#1}} 

\author[1,3]{Gloria Fabris}
\author[2]{Alessandro Lucantonio}
\author[1]{Nico Hampe}
\author[1]{Erik Noetzel}
\author[1]{Bernd Hoffmann}
\author[1]{Antonio DeSimone}
\author[1]{Rudolf Merkel\thanks{Corresponding author -- e-mail address:\,\texttt{r.merkel@fz-juelich.de}.}}
\affil[1]{\small{Institute of Complex Systems, ICS-7: Biomechanics, Forschungszentrum Jülich, Jülich, Germany.}}
\affil[2]{\small{SISSA--International School for Advanced Studies, via Bonomea 265, 34136 Trieste, Italy.}}
\affil[3]{\small{Department of Mechanical Engineering, Stevens Institute of Technology, Hoboken, NJ, USA.}}

\title{Nanoscale topography and poroelastic properties of model tissue breast gland basement membranes}

\date{}

\begin{document}

\maketitle
	
\begin{abstract}
Basement membranes (BMs) are thin layers of condensed extracellular matrix proteins serving as permeability filters, cellular anchoring sites, and barriers against cancer cell invasion. It is believed that their biomechanical properties play a crucial role in determining cellular behavior and response, especially in mechanically active tissues like breast glands. In spite of this, so far relatively little attention has been dedicated to their analysis due to the difficulty of isolating and handling such thin layers of material. Here, we isolated basement membranes derived from MCF10A spheroids - 3D breast glands model systems mimicking {\it in-vitro} the most relevant phenotypic characteristics of human breast lobules - and characterized them by atomic force microscopy (AFM), enhanced resolution confocal microscopy (LSM), and scanning electron microscopy (SEM). By performing AFM height-clamp experiments, we obtained force relaxation curves that offered the first biomechanical data on isolated breast gland BMs. Based on LSM and SEM imaging data, we modeled the system as a polymer network immersed in liquid and described it as a poroelastic material. Finite Element (FE)  simulations matching the experimental force relaxation curves allowed  for the first quantification of the bulk and shear moduli of the membrane, as well as its water permeability. These results represent a first step towards a deeper understanding of the mechanism of tensional homeostasis regulating mammary gland activity, as well as its disruption during processes of membrane breaching and metastatic invasion.
\end{abstract}

\section*{Introduction}
Basement membranes (BMs) are thin sheets of condensed extra-cellular matrix proteins secreted by epithelial, mesothelial and endothelial tissues that separate them from underlying connective tissue \cite{alberts}. They are ubiquitously present in the body and provide a variety of functions \cite{ECM_interaction}: besides acting as cellular anchoring sites, they regulate cellular motility, influence tissue remodeling, and, in some cases, even act as highly selective permeability barriers \cite{plos_10, paper_Aljona}. Their composition varies according to physiologic or pathologic conditions \cite{bm_structure} and to tissue distribution, but their main components are collagen (typically, type IV), various laminin isoforms, nidogen, and perlecan. 
Typically, laminins start the process of BM assembly by binding to cell surface receptors, most prominently of the $\beta$1 integrin family. Subsequently, self-assembly of laminin and attraction to other BM components occur. Nidogens act as network stabilizers by bridging laminins and type IV collagen, which polymerizes forming a covalently linked network \cite{Coll_1, Quonda}. Perlecan and agrin then bind to nidogen, laminin, integrins, dystroglycan, and sulfated glycolipids, creating multiple collateral interactions whose exact molecular details are still the object of intense scrutiny \cite{Yuri}.

In breast glands, BMs are fundamental components of epithelial tissue architecture and act as the first barrier against metastatic invasion \cite{bm_cancer_1}; the process of membrane breaching initiated by invasive neoplasms is partly mediated by the secretion of BM-specific matrix metalloproteinases (MMPs), and further facilitated by the increased motility and proliferative potential of cancer cells \cite{BM_invasion}. The alterations in tensional homeostasis caused by expanding tumor masses can in fact lead, over time, to a break-through of invasive cells into the neighboring connective tissue \cite{tense_sit}. 
In order to better comprehend the mechanisms of membrane breaching during cancer cell invasion, therefore, specific BMs' biophysical properties (such as thickness, stiffness and permeability) ought to be characterized. Despite the high physiological relevance of breast gland BMs, however, no single study has tried to biomechanically characterize them yet due to the difficulty of handling such thin protein layers.

Here, basement membranes were investigated in a simplified environment with respect to the {\it in-vivo} situation. We isolated and characterized BMs endogenously secreted by 3D cellular spheroids (or \textit{acini}) derived from the human breast epithelial cell line MCF10A \cite{debnath}. Such 3D cell cultures recapitulate, {\it in-vitro}, the most relevant physiological features of breast gland acini {\it in-vivo}, and are therefore a precious platform to investigate the fundamental units of breast gland tissue in a biologically relevant and yet controlled context. 
Our previous work highlighted that the BM scaffolds of MCF10A spheroids develop gradually and can be categorized as low - (1 to 12 days in culture), semi-  (13 to 24 days), and highly- (more than 24 days) matured \cite{paper_Aljona}. In the present work, BMs isolated from low- and highly- matured acini were analyzed and compared. 

Some basement membranes isolated from other types of tissues have already been the object of similar topographical and biomechanical studies: SEM imaging, for instance, has been used to estimate the pore size of vascular endothelial BMs \cite{macaco}. AFM imaging and indentation, on the other hand, have revealed that the thickness and elasticity of the retinal internal limiting membrane (ILM) are age-dependent \cite{bm4}, change dramatically according to hydration level \cite{bm_H2O}, and display a side-specificity \cite{bm3,bm_bello} as a consequence of asymmetric protein organization. Analogous results have been reported for corneal BMs \cite{cornea}. 
The approach used so far to extract values of the Young's modulus from AFM indentation data (namely, the use of the Hertz model), however, assumes that the material analyzed is a homogeneous, half-infinite, and perfectly elastic space \cite{Hertz}. 
Clearly, when indenting very thin layers of hydrated biopolymer networks immersed in water, this approximation cannot hold. 

Here, based on the indications on the BM structure obtained from LSM and SEM imaging, we modeled the membrane as a polymer network immersed in liquid. We developed a large deformation poroelastic model which, by accounting for the finite thickness of the membrane, describes its mechanics under deep indentation and allows to estimate both mechanical properties and water permeability of the BMs. In short, the present study offers the first characterization of breast gland basement membranes in terms of topography, structure and mechanical properties, and brings us one step closer to a deeper understanding of epithelial tissue architecture and regulation under normal and pathological conditions.

\section*{Materials and Methods}
\subsection*{Cell culture}
MCF10A cells (ATCC, Manassas, VA USA) were cultured in a humidified environment (5\% CO$_2$, $37^\circ$C) in DMEM/ F12 growth medium (Life Technologies, Darmstadt, Germany) containing 5\% horse serum (Life Technologies, Darmstadt, Germany), 20~ng/mL epidermal growth factor (EGF, Sigma-Aldrich, Steinheim, Germany), 0.5~\si{\micro\gram}/mL hydrocortisone (Sigma-Aldrich, Steinheim, Germany), 100~ng/mL cholera-toxin (Sigma-Aldrich, Steinheim, Germany), 10~\si{\micro\gram}/mL insulin (Sigma-Aldrich, Steinheim, Germany), 100~U/mL penicillin (Life Technologies, Darmstadt, Germany), and 100~\si{\micro\gram}/mL streptomycin (Sigma-Aldrich, Steinheim, Germany). MCF10A 3D acini were cultivated according to a protocol adapted from Debnath et al.~\cite{debnath} and described in \cite{paper_Aljona}. In short, single cells were seeded on a growth factor-reduced EHS gel bed (Geltrex, Life Technologies, Darmstadt, Germany) and supplemented with DMEM/F12 assay medium containing 2\% horse serum, 5~ng/mL EGF (exclusively day 1 to 9), 0.5~\si{\micro\gram}/mL hydrocortisone, 100~ng/mL cholera toxin, 10~\si{\micro\gram}/mL insulin, 100~U/mL penicillin, 100~\si{\micro\gram}/mL streptomycin and a low, non-gelling concentration of Geltrex (2\%), which was changed every third day.
 
\subsection*{Basement membrane isolation}	
\label{BM_isolation}
\begin{figure*}[!t]
	\centering
	\includegraphics[width=14.5cm]{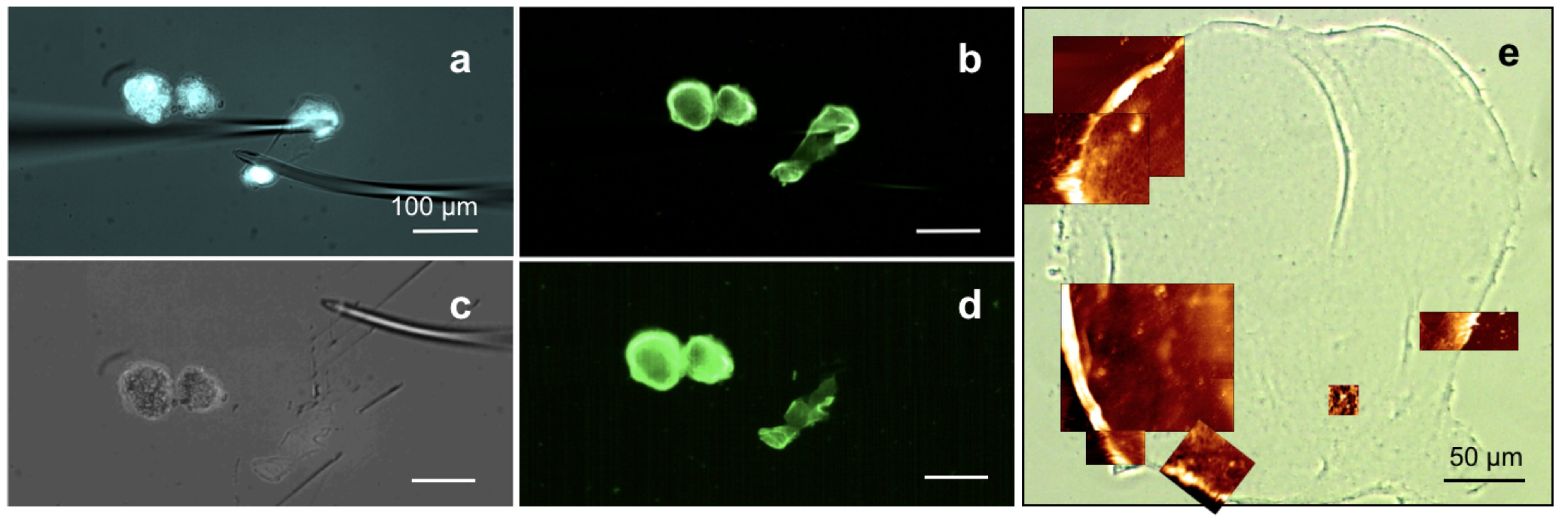}
	\caption{ (a) Isolation procedure for highly-matured BMs. MCF10A spheres were isolated from the EHS matrix and transferred on a poly-L-lysine treated Petri dish, then the membrane was peeled off breaking the spheres apart with bent microcapillaries. (a) In cyan: CellTracker cytoplasmic staining. (b) In green: collagen IV staining. (c) Phase contrast image of the isolated BM fragments and (d) corresponding collagen IV staining. Only parts of the membrane lying flat and not folded on the surface were used for AFM imaging. (e) Phase contrast light microscopy image of a low-matured isolated basement membrane in ultrapure water. In brown: overlapping AFM topographical images recorded in contact mode.}
	\label{fig:BM_peel}
\end{figure*}

First, acini had to be isolated from the EHS matrix: to this end, samples were first washed with ice cold phosphate buffered saline (PBS, 5 min), then incubated in cell recovery solution (CRS, BD Bioscience, Fernwald, Germany) for 45 min at 4$^\circ$C. Individual spheres were carefully pipetted out of the fluid gel matrix under a stereo microscope (Stemi 2000-CS, Carl Zeiss, Jena, Germany) and transferred in centrifuge tubes treated for minimizing protein binding (LoBind, Eppendorf, Wesseling-Berzdorf, Germany). After centrifugation (5 min, 0.1 RCF, 4$^\circ$C) and elimination of the supernatant, a pellet of MCF10A acini could be resuspended and transferred on poly-L-lysine-coated glass (30 min, 37$^\circ$C) for further staining and imaging, or kept for BM isolation. For AFM and SEM characterization, two different isolation protocols were adopted depending on the acinar maturation stage. For MCF10A acini up to 12-15 days, after the isolation from the Geltrex matrix it sufficed to perform one extra round of centrifugation at 16 RCF (4 min, 4$^\circ$C). The acinar structures then literally broke apart, leaving thin BM fragments floating in the supernatant. These could then be carefully pipetted on poly-L-lysine-coated glass coverslips (30 min, 37$^\circ$C), where they would adhere firmly. For older spheres, centrifugation was not sufficient for isolating the BM. A manual setup for membrane isolation had to be established: this consisted of two self-produced bent glass microcapillaries that, being controlled via separate micromanipulators, could be employed to immobilize the MCF10A acinus on one end, and break through it on the other end, effectively allowing a partial peeling off of BM fragments (see Fig.~\ref{fig:BM_peel}a-d). For BM visualization, a collagen IV staining was performed prior to isolation (see Fig.~\ref{fig:BM_peel}b,d). In both cases, after BM isolation, the samples were treated with a 1\% solution of octyl-$\beta$-D-glucopyranoside detergent (Sigma-Aldrich, Steinheim, Germany) in PBS and subject to ultrasonic treatment (Sonorex RK-100, Bandelin, Berlin, Germany) for 15-20 min in order to eliminate cellular debris and lipid residues from the membranes. 

\subsection*{Immunocytochemistry and enhanced resolution confocal microscopy}
For immunocytochemistry of the BMs of native MCF10A acini, these were isolated from the EHS gel matrix as described above and transferred on poly-L-lysine treated glass coverslips. Unspecific antibody binding was blocked by incubation with 5\% skim milk powder in PBS for 30 min at 37\si{\degree}C. The samples were not fixed in order to image BMs under native conditions. 1:200 solutions of primary antibodies binding to collagen IV (rabbit, ab6586, Abcam, Cambridge, ENG, UK), collagens I-II-III (clone MMCHABC, mouse, MAB1334, Merck, Darmstadt, Germany), laminin-3A32 (gamma-2 chain, clone D4B5, isotype IgG1, mouse, MAB 19562, Merck, Darmstadt, Germany) and perlecan (anti-heparan sulfate proteoglycan antibody, clone A7L6, rat, MAB1948P, Merck, Darmstadt, Germany) were incubated in dilution buffer (1\% skim milk powder in PBS) for 2 h at 37\si{\degree}C. After a washing step in dilution buffer (5 min, RT), incubation with the secondary antibodies solution composed of 1:200 anti-rabbit IgG (Alexa-Fluor 405, A31556, ThermoFisher, Waltham, MA, USA), 1:1000 goat anti-mouse IgM (Alexa Fluor 488, A21042, ThermoFisher, Waltham, MA, USA), 1:200 anti-mouse IgG1 (Alexa-Fluor 647, A21240, ThermoFisher, Waltham, MA, USA) and 1:200 anti-rat (eFluor 570, 41-4321-82, eBioscience, Darmstadt, Germany) antibodies in dilution buffer followed (45 min, 37\si{\degree}C). After gentle aspiration of the solution and a washing step in PBS (5 min, RT), the sample was embedded in FluoroMount (Sigma-Aldrich, Steinheim, Germany) to be analyzed. Confocal microscopy was performed using an LSM 880 with Airyscan detector allowing for lateral resolutions down to 130 nm (Carl Zeiss, Jena, Germany). In this setup, the complete Airy disk is imaged on a concentrically arranged array of hexagonal detectors consisting of 32 elements of 0.2 AU each. During acquisition, the pinhole remains open hence maximizing signal collection. Image reconstruction with improved resolution is obtained via pixel reassignment of the signals from all detector elements to their correct position and partial deconvolution \cite{AiryScan}.

\subsection*{Collagen IV pore size and filament thickness determination}
\noindent For determining collagen IV pore sizes, fluorescent images were filtered using a 2D Gaussian filter with standard deviation of 0.5~pixels. Then the image background was identified by morphological opening using a disk-shaped structuring element with a radius of 5~pixels (pixel size: 0.042~\si{\micro\meter}). In a next step, the background image was subtracted from the smoothed image and the result was again smoothed with a Gaussian filter of standard deviation value of 3 pixels. After contrast enhancement, the image was segmented using half of its mean gray value as a threshold. Everything below the threshold was consequently labeled as 'hole'. At least $n=10$ images of different BMs were analyzed for each population.
For determining collagen filaments' thickness, fluorescent images were preprocessed with the same procedure used in the pore size finding algorithm. After contrast enhancement and image segmentation, everything above the threshold was labeled as 'filament'. Filaments with less than 100~pixels were rejected (pixel size: 0.042~\si{\micro\meter}). Then, the distance of each pixel belonging to the skeletonized filament mask to the nearest black pixel was calculated, thus identifying the local filament radius. Again, at least $n=10$ images of different BMs were analyzed for each population.

\subsection*{Atomic force microscopy}
All AFM measurements were performed using a Nanowizard Life Science version instrument (JPK, Berlin, Germany) equipped with an inverted optical microscope (Axiovert 200, Carl Zeiss, Jena, Germany) for sample observation. For imaging, pyramidal silicon nitride tips with nominal spring constant $k=0.06$~N/m and resonance frequency $f=18$~kHz (DNP-10 D, Brucker, Leiderdorp, The Netherlands) were used in contact mode. Typical scan areas ranged from 15$\times$15~\si{\micro\meter^2} to 50$\times$50 \si{\micro\meter^2} (see Fig.~\ref{fig:BM_peel}e). For force relaxation experiments, tips with indenters of two different radii were used: a silicon tip having a nominal $k=0.2$~N/m and $f=13$~kHz and terminating in a spherical indenter of $R=500$ nm (B500-CONTR, Nanotools, Munich, Germany) and a silicon tipless cantilever of nominal $k=0.04$~N/m and $f=7$~kHz (Arrow TL1Au with Ti/Au back tip coating, Nanoworld, Neuchatel, Switzerland) modified by the attachment of a $R=3.5$~\si{\micro\meter} silica bead (PSI-5.0, surface plain, G. Kisker GbR, Steinfurt, Germany) via two-component glue (plus Endfest 300, UHU, Buehl Baden, Germany). For each BM, force relaxation data were typically recorded on an 8x8 grid of positions and repeated varying the force setpoint for both indenters. Data were averaged over a subset of the grid (after removing curves affected by instrument drift, instability or similar artifacts) to account for local structural heterogeneities.
All tips were individually calibrated via the thermal noise method \cite{afm_calib}. The cantilever speed was held at $v=5$~\si{\micro\meter}/s, and the force setpoint varied from $0.1$~nN to $1$~nN. 

\subsection*{Scanning electron microscopy}
For SEM experiments, isolated BMs on glass cover slides were additionally washed in a 0.5\% solution of Triton-X-100 (Sigma-Aldrich, Steinheim, Germany) in PBS (5 min, RT) before ultrasonic treatment in PBS (10 min, RT). After dehydration in a graded series of HPLC-grade ethanol in ultra-pure water (10\%, 30\%, 50\%, 70\%, 90\%, 95\% (3x), 100\%, 5 min, RT) and critical point drying in CO$_2$ (CPD 030, Bal-Tec, Balzers, Liechtenstein), a 2~nm coating of Pt/Pd was sputtered on the samples using a 208 HR sputter coater (Cressington, Watford, ENG, UK) with MTM-20 thickness controller unit. Imaging was performed on a Gemini 500 SEM (Carl Zeiss, Jena, Germany) using an acceleration voltage between 3 and 10~kV at magnifications ranging from 8,500$\times$ to 265,000$\times$.

\subsection*{Computational modeling of the indentation experiments} 
Finite element simulations of the indentation experiments were performed using a large-deformation poroelastic model \cite{Ale} to characterize the coupled elasticity and fluid transport of the basement membranes.
In such a model, the state of the membrane is described by the displacement field $\mb{u}$ of the polymer network with respect to the reference configuration and the solvent concentration $c$ per unit reference volume. The chemical potential $\mu$ of the solvent (here, water) within the membrane quantifies the energy carried by the solvent and its gradient represents the driving force of solvent migration. We will use the symbol $\mb{F}=\mb{I}+\nabla\mb{u}$ (with $\mb{I}$ the identity) for the deformation gradient and write $J=\det\mb{F}$ for its determinant. Swelling processes are governed by the equations of balance of forces and moments that, assuming inertia negligible, read:
\begin{align}
\label{eq:balaforces}
\divg \mb{S} = \mb{0}\,, \qquad \mbox{skw}\,\mb{S}\mb{F}^{\rm T} = \mb{0}\,,
\end{align}
where $\mb{S}$ denotes the first Piola-Kirchhoff stress tensor. By the balance of solvent mass:
\begin{align}
\label{eq:balasolgel}
\dot{c} = \divg \left(\frac{c D}{\mc{R} T} \nabla \mu\right)\,,
\end{align}
where $D$ is the water diffusivity, $c$ the concentration of water per unit volume of the undeformed membrane, $\mc{R}$ the universal gas constant and $T$ the absolute temperature.
Here, a Darcy-like law was employed to relate the solvent flux to the gradient of solvent chemical potential because the polymer matrix and the solvent are considered to be separately incompressible; hence, the change in volume of the membrane is related to the change in solvent concentration with respect to its initial value $c_{\rm o}$:
\begin{align}
\label{eq:swellconstr}
J = 1 + \Omega (c-c_{\textrm{o}})\,.
\end{align}
This constraint is enforced through the Lagrange multiplier $p$. As concerns the constitutive equations, we prescribe the following Flory-Rehner representation for the free energy density of the membrane \cite{Flory1,Flory2}:
\begin{align}
\psi(\mb{F},c) = \psi_{\rm e}(\mb{F}) + \psi_{\rm m}(c) \,,
\end{align}
where $\psi_{\rm e}(\mb{F})$ and $\psi_{\rm m}(c)$ are the neo-Hookean elastic energy of the polymer network and the Flory-Huggins free energy of solvent-polymer mixing, respectively. The parameters of the free energy density are: the shear modulus $G_{\textrm{d}}$ of the dry polymer, the polymer-solvent mixing parameter  $\chi$ and the absolute temperature $T$ of the environment. Consistency with thermodynamical principles provides the constitutive equations for $\mb{S}$ and $\mu$ as the derivatives of the free energy with respect to $\mb{F}$ and $c$, respectively (with the additional reactive terms depending on the Lagrange multiplier $p$). By linearizing the constitutive equations for the stress about the initial configuration, the incremental shear $G$ and bulk $K$ moduli of the membrane can be computed as \cite{Lucantonio2012}:
\begin{align}
\label{eq:bulk}
G = \frac{G_{\rm d}}{J_{\rm o}^{1/3}}\,, \quad K = -\frac{G}{3}+\frac{\mc R T}{\Omega} \left(\frac{1}{\phi_{\rm o}}-2\chi \right)\left(1-\phi_{\rm o}\right)^2\,,
\end{align}
where $\phi_{\rm o}=1-1/J_{\rm o}$ is the initial volume fraction of water (unindented condition), which is related to the initial water concentration: $c_{\textrm{o}} = (J_{\textrm{o}}-1)/(\Omega J_{\textrm{o}})$. In the reference configuration, since the membrane is stress-free and in contact with water in equilibrium with its vapor, the following mechano-chemical equilibrium condition \cite{Lucantonio2012} holds:
\begin{align}
\label{eq:chemeq}
\frac{\mc R T}{\Omega}\left[\log{\phi_{\rm o}}+1-\phi_{\rm o} + \chi (1-\phi_{\rm o})^2 \right] + G = 0\,.
\end{align}
In particular, once $\phi_{\rm o}$ and $\chi$ are known, the incremental shear modulus $G$ can be readily computed from Eq.~\ref{eq:chemeq}. Hence, the model depends on three independent poroelastic material parameters: $D$ and two others among those appearing in Eqs.~\ref{eq:bulk}-\ref{eq:chemeq}, namely $\phi_0$, $\chi$, $G$, $G_d$, and $K$. The poroelastic parameters were fitted numerically for each membrane in order to reproduce the experimental force-relaxation curves obtained using indenters of radius $R=3.5$~\si{\micro\meter} and $R=0.5$~\si{\micro\meter}, for the AFM force setpoints $F_{\rm max} = 1$~nN and $F_{\rm max} = 0.5$~nN, respectively. Specifically, the gradient-free BOBYQA optimization algorithm \cite{Powell2009} was employed to perform a least-square fitting of the experimental data. For symmetry reasons, 1/4 of each membrane was chosen as the computational domain. It was numerically verified that, due to the high ratio between the in-plane dimensions and the thickness measured in the experiments, the membranes behave mechanically as solids with infinite planar extension, so that the specific geometry of the boundary is irrelevant. Thus, each membrane was modeled as a parallelepiped with a sufficiently high aspect ratio (100), and a thickness set to the individual average value measured experimentally via AFM imaging. The substrate was assumed to be rigid, frictionless and impermeable. Apart from the boundaries in contact with the substrate and the indenter, the free surface of the membrane was assumed to be in chemical equilibrium with the surrounding water bath at all times, a condition that amounts to prescribing a null chemical potential (\textit{i.e.}~pure water in equilibrium with its own vapor) on the wet surface. A penalty formulation was employed to describe frictionless contact between the membrane and the spherical indenter, which was modeled as a rigid body. The time-history of the indenter displacement was prescribed according to the experiments. 
Detailed information on the computational model and the numerical procedures, including a table of the parameter values used in the simulations, can be found in the Supporting Materials (SM 1, Table S1).

\section*{Results}

\begin{figure*}[!t]
	\centering
	\includegraphics[width=16.5cm]{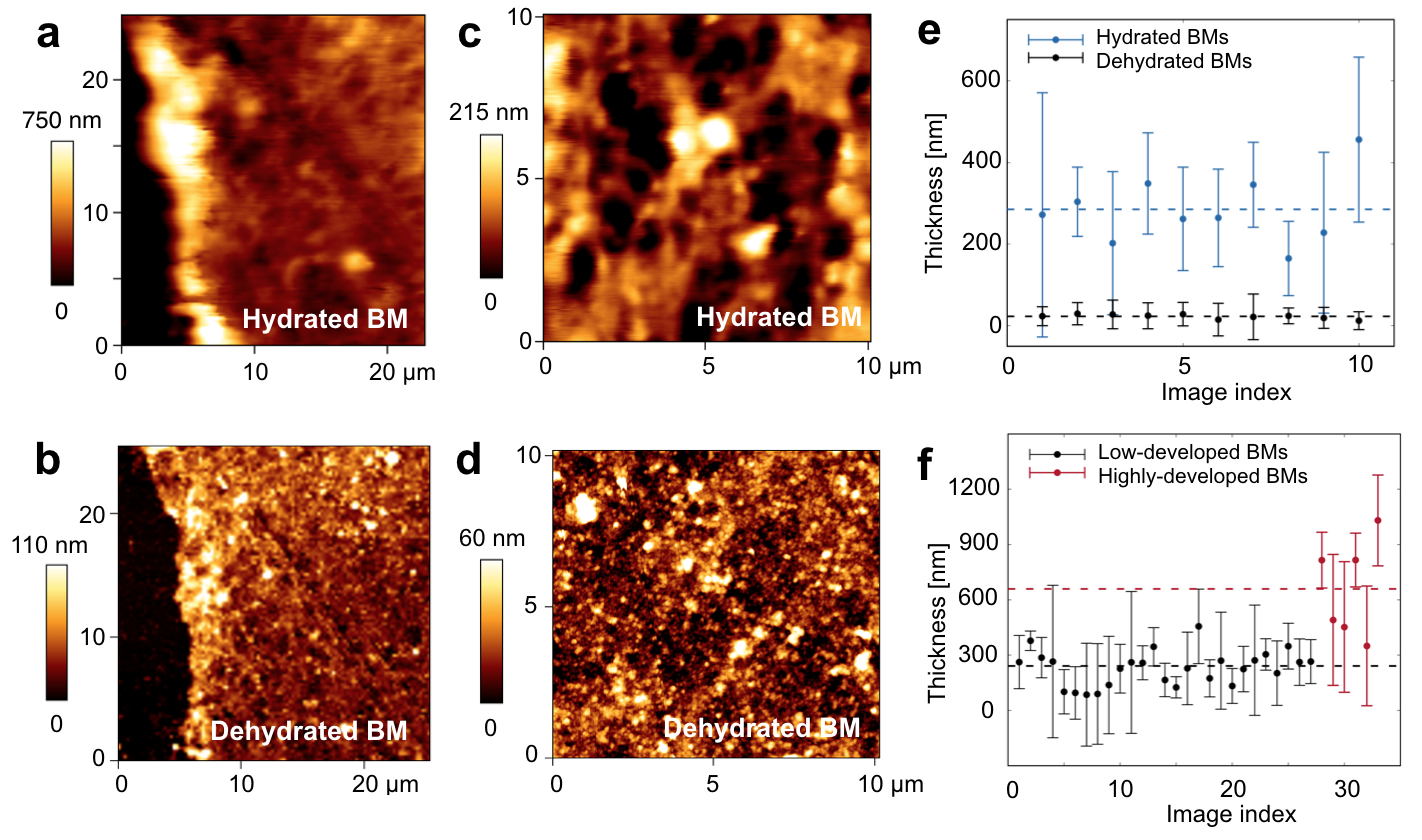}
	\caption{(a) Representative contact-mode imaging of a portion of BM isolated from a low-matured (day 12) BM in ultrapure water and (b) imaging of the same portion in air, after dehydration. The black background is the glass coverslip. (c) Surface topography of an inner portion of a hydrated and (d) dehydrated BM, respectively. (e) Decrease in mean BM thickness of low-matured membranes following dehydration and (f) age-dependent increase in BM thickness between low- and highly-matured MCF10A acini. The distributions represent mean and FWHM of the gaussian distributions used for fitting the height histograms of single AFM images. The horizontal dotted lines represent average values of the distributions centers.}
	\label{fig:AFM_1}
\end{figure*}

\subsection*{AFM topographic characterization shows variations in BM thickness during membrane maturation and dehydration}
In order to characterize the topography and thickness of basement membranes during the various maturation stages, BMs were isolated from low- and highly-matured MCF10A acini and assessed by means of AFM contact imaging in ultrapure water. As expected for such a complex biological material, a certain level of heterogeneity was observed among the samples.
For low-matured spheres (day 6-12), 27 images recorded from 17 different membranes were analyzed in order to obtain histograms of pixel heights; given the increased difficulty of isolating highly developed BMs, for highly matured ones only six images from three different membranes (day 28-31) could be obtained. Despite the relatively low number of samples analyzed, a clear difference could be measured between the thickness of membranes at different stages of maturation (see Fig.~\ref{fig:AFM_1}f). Single height histograms could always be fitted by Gaussian distributions, and averaging over the distribution centers $x_c$, mean thickness values of 230 nm (with s.d. 95~nm) and 660 nm (s.d. 265~nm) were obtained for low- and highly- matured BMs, respectively. This trend is well in line with data reported for other BMs \cite{bm_H2O} and reveals a thickening of the basement membrane caused by endogenous protein secretion, as confirmed by immunostainings performed with anti-human laminin-3A32 antibodies.

Upon sample dehydration, a tenfold decrease in membrane thickness was observed (see Fig.~\ref{fig:AFM_1}e), with mean thickness values as low as 22~nm ($n=10$, s.d. 6~nm). This result underlines the crucial role played by water in conferring BMs their native structure and confirms the necessity of not merely modeling this material as an elastic solid, but rather as a hydrated matrix immersed in fluid.

\begin{figure*}[!t]
	\centering
	\includegraphics[width=12.5cm]{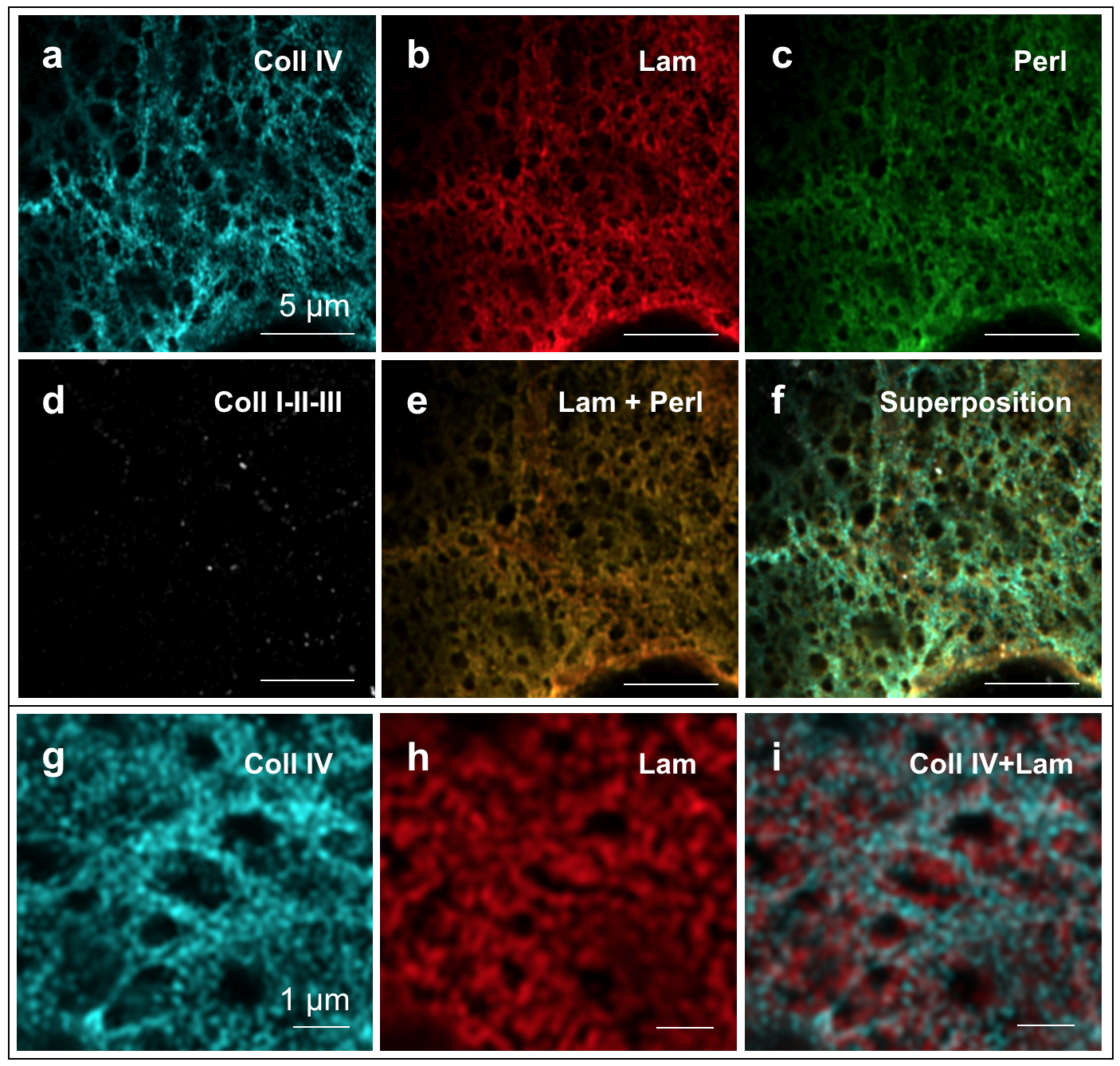}
	\caption{Immunocytochemical stainings of (a) type IV collagen, (b) laminin-3A32, (c) perlecan and (d) interstitial collagens (type I-II-III). (e) Colocalization of laminin-3A32 and perlecan signals (b+c). (f) Merged image (a+b+c+d). Zooming in (g) the collagen IV rings and (h) the laminin-3A32 structure, it becomes apparent how these create two intertwined meshworks of different characteristic pore size. (i) Merged signals (g+h). Scale bars: 5~\si{\micro\meter} (a-f), 1~\si{\micro\meter} (g-i).}
	\label{fig:LSM}
\end{figure*}

\subsection*{LSM enhanced resolution imaging of main BM components show collagen IV- dependent architecture}
Immunostainings of the main components of breast gland BMs were performed on whole MCF10A acini of different maturation states isolated from the EHS gel and transferred on glass coverslips. Enhanced resolution confocal microscopy with Airyscan detector was used to image collagen type IV, laminin-3A32, the proteoglycan perlecan, and collagens types I-II-III within the BM. 
 
\begin{figure*}[!t]
	\centering
	\includegraphics[width=12.5cm]{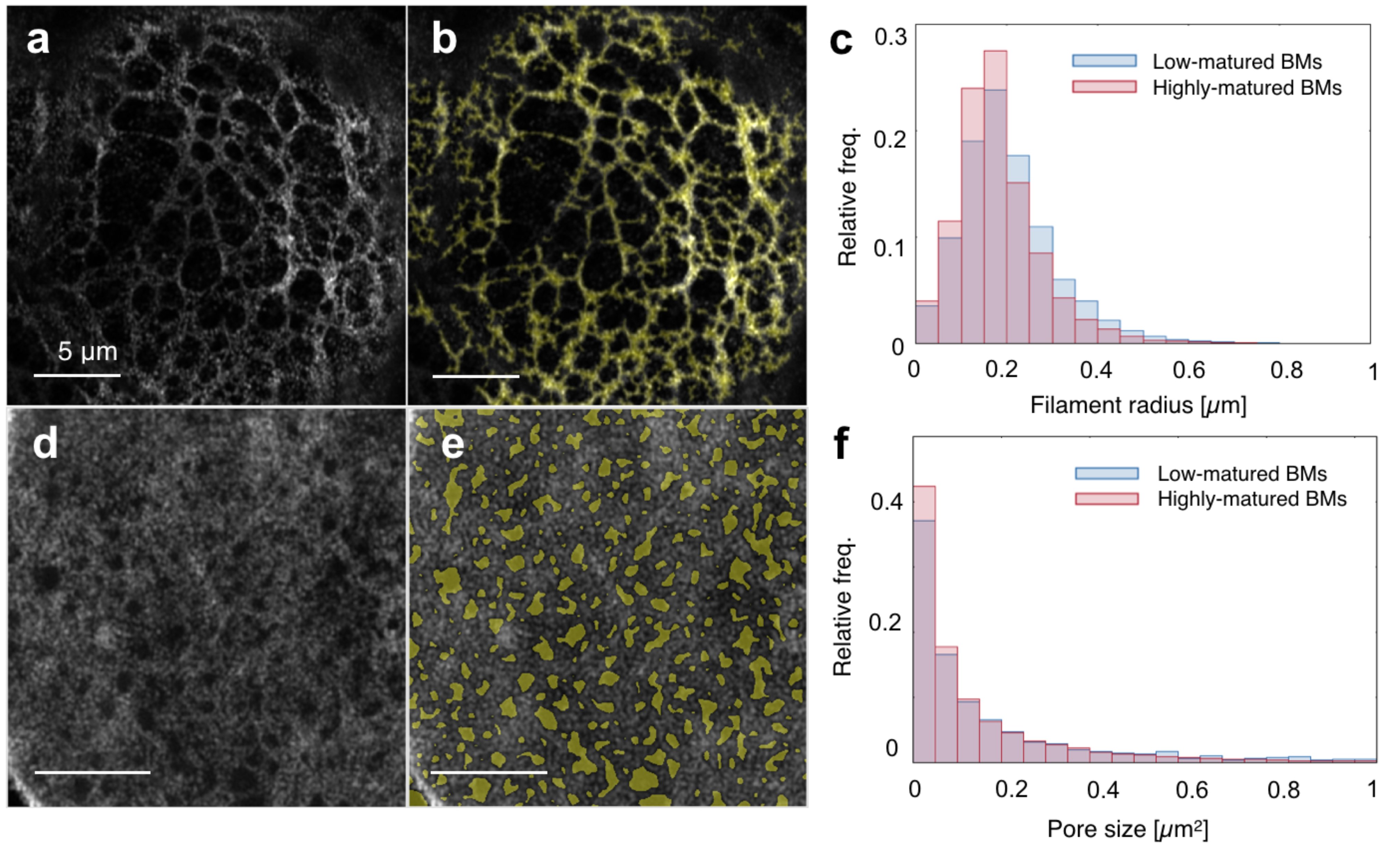}
	\caption{(a) Airy scan fluorescent image of the collagen IV network. (b) Filament mask. (c) Distribution of filament radii for low- and highly-matured basement membranes as determined from the masks. (d) Airy scan fluorescent image of the collagen IV network. (e) Pores mask. (f) Corresponding pore size distributions. All scale bars: 5~\si{\micro\meter}.}
	\label{fig:pores}
\end{figure*}

The stainings revealed a clear structure in protein architecture (see Fig.~\ref{fig:LSM}). A principal network composed of collagen IV was observed in BMs at all developmental stages. This was organized in the typical meshwork configuration ascribed to non-fibrillar collagens \cite{fratzl} and had a variable pore size spanning from about 50~nm$^2$ up to almost 1~\si{\micro\meter^2} (see Fig.~\ref{fig:LSM}a, g and Fig.~\ref{fig:pores}f). No significant difference could be observed in the pore diameter distribution between low-matured and highly-matured BMs (see Fig.~\ref{fig:pores}f); similarly, also the visible collagen filament thickness remained essentially unchanged during BM development, with an average filament radius of about 200~nm (see Fig.~\ref{fig:pores}c). With the caveat that the resolution limit of Airy scan microscopy (in this case approximately 30\% of the excitation wavelength) prevented us from observing molecular structures, this implies an early organization of collagen IV into its final structural arrangement.  

Laminin-3A32, on the other hand, does not form a covalently linked independent network \cite{pozzi}, but was arranged in a denser structure of smaller characteristic size that interconnected with the collagen IV meshwork often partially filling its pores (see Fig.~\ref{fig:LSM}i). Even after optimization of the filter settings in order to prevent bleed-through, the perlecan signal was practically indistinguishable from that of laminin-3A32, indicating a high degree of co-localization between the two proteins for membranes of both age groups (see Fig.~\ref{fig:LSM}e). The absence of interstitial collagens was confirmed by immunostainings recognizing specifically collagens of type I-II-III (see Fig.~\ref{fig:LSM}d). 

\subsection*{SEM images reveal two levels of protein organization}
\begin{figure*}[!t]
	\centering
	\includegraphics[width=16.5cm]{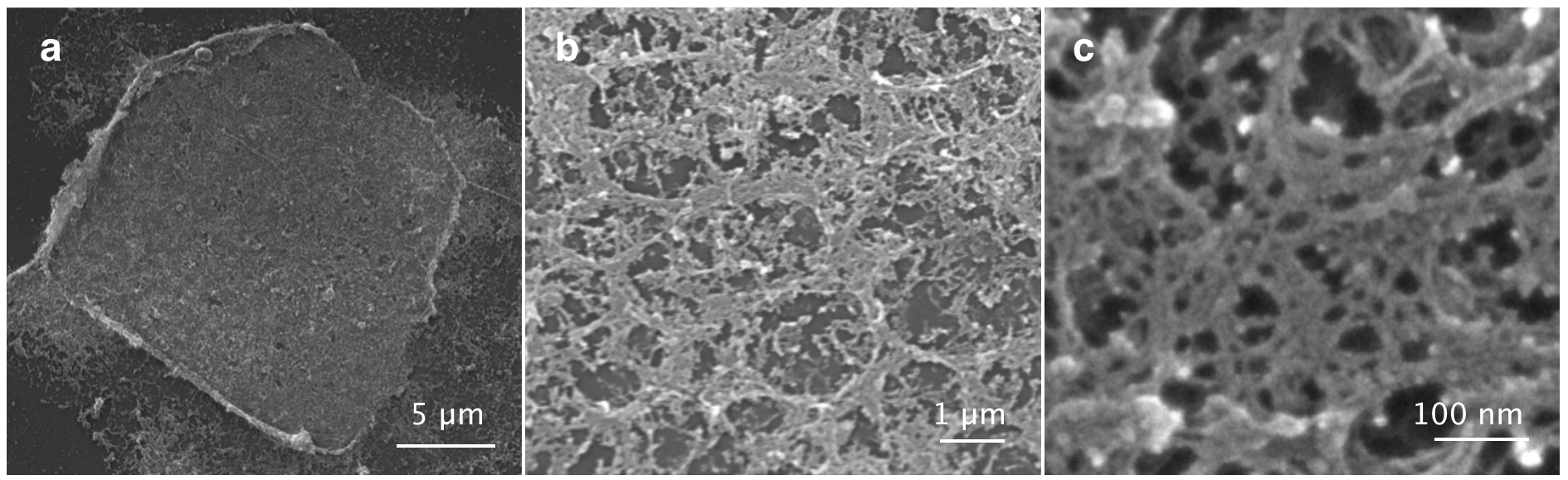}
	\caption{(a) SEM micrograph displaying the overview of a low-matured (day 8) isolated BM lying flat on the substrate. Despite the many fractures present in the BM as a consequence of sample preparation, in (b) a series of regular, ring-like bundles of fibers constituting the backbone of the BM can be identified, while (c) shows zoom-ins on the finer protein structure composing the BM meshwork.}
	\label{fig:SEM}
\end{figure*}

For further BM ultrastructure characterization, isolated membranes were observed via scanning electron microscopy (see Fig.~\ref{fig:SEM}).
This time, because of the difficulty of isolating highly-matured BMs, only membranes from the low-matured group (day 6-12) were analyzed. Despite critical point drying, following SEM preparation numerous fractures and holes appeared on the membranes, indicating damages to the finer protein structures composing the BM. 
This inconvenience, though, allowed us to identify a 'backbone' that remained intact in all samples (see Fig.~\ref{fig:SEM}b). This appeared as a meshwork of thick protein bundles arranged in a honeycomb-like structure with a characteristic size of about 1~\si{\micro\meter} in diameter. Even accounting for a shrinking of about 20\% in the volume of biological samples treated for SEM preparation, this mesh size is very well in line with the one observed for the collagen IV pores in the LSM images. Within this polymer network, a finer meshwork of thin fibers could be observed (see Fig.~\ref{fig:SEM}c).
Even though SEM data does not allow for discrimination between different structural components of the BM, the high resolution of the images gives us a clear idea of the two-level organization of breast gland basement membranes.

\begin{figure*}[!ht]
	\centering
	\includegraphics[scale=1]{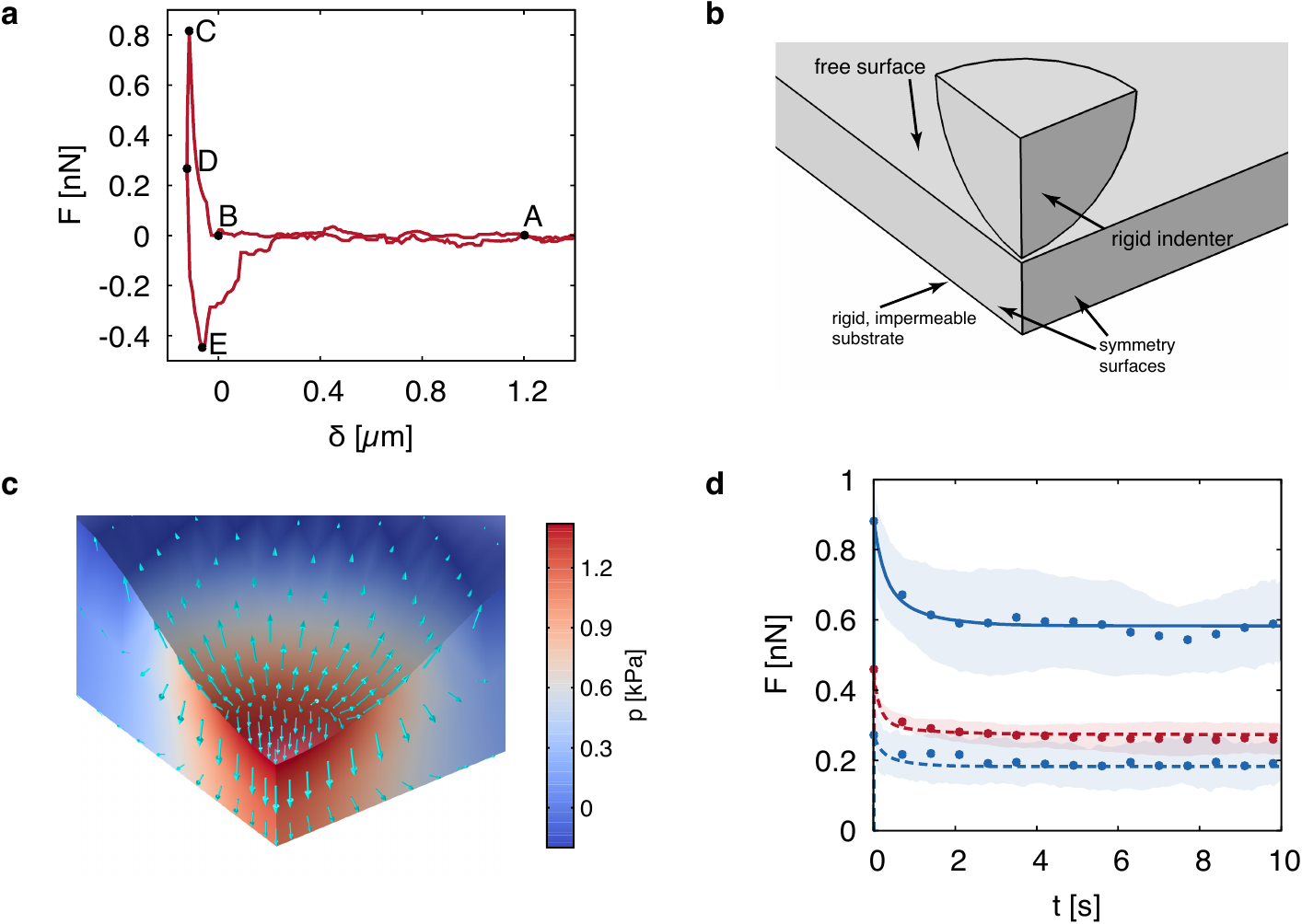}
	\caption{(a) Typical experimental force-indentation curve. In each experimental run, the indenter is displaced from the initial position (A) past the contact point with the membrane surface (B, $\delta=0$ \si{\micro\meter}), until approaching the force setpoint $F$ (C). Then, the base of the AFM cantilever is held at a fixed position, while the tip moderately sinks into the sample, as the force decreases down to a steady value (D) across a time period of 10-15~s. Upon indenter retraction, the force reaches a minimum (E) corresponding to the maximum adhesion force between the indenter and the membrane. (b) Schematic of the computational domain for the finite element simulations. (c) Snapshot of the deformed configuration of the membrane at peak indentation force (indenter not shown), as computed from numerical simulations. The region shown corresponds to 1/4 of the membrane. The color code represents the solvent pressure field $p$.  Arrows represent the water flux vector (Max: $4.5 \times 10^{-2}$~\si{\mole/\meter\squared\s}. Logarithmic scale). (d) Force-relaxation curve: comparison between experimental (dots) and numerical (lines) results for the membrane BM2, for different values of the indenter radius $R$: $3.5$~\si{\micro\meter} (blue), $0.5$~\si{\micro\meter} (red), and force setpoint $F$: $1$~nN (solid lines), $0.5$~nN (dashed lines). Dots and shaded areas correspond to the spatial averages (i.e. averages over all measured positions for a certain BM) and the standard deviations of the indentation force as computed from experimental data, respectively.}
	\label{fig:simulations}
\end{figure*}

\subsection*{BM mechanical properties characterized by AFM showed behavior typical of poroelastic materials}
To probe the mechanical and permeation properties of breast gland basement membranes, AFM indentations in height-clamp mode using two different spherical indenters (of radius $R=0.5$~\si{\micro\meter} and $R=3.5$~\si{\micro\meter}, respectively) were performed on low-matured BMs ($n=5$) isolated from MCF10A acini. A typical experimental indentation force curve is shown in Fig.~\ref{fig:simulations}a. For each BM, force relaxation data were typically recorded on an 8x8 grid of positions and repeated varying the force setpoint for both indenters. Data were averaged over a subset of the grid (after removing outliers, \textit{i.e.}~curves affected by excessive drift or obvious instabilities) to account for local structural heterogeneities.

\begin{table}[!b]
	\centering
	\begin{tabularx}{0.81\textwidth}{ l c c c c c c c c}
		\toprule
		BM\# & \multicolumn{2}{c}{$G$ (Pa)} & \multicolumn{2}{c}{$K$ (Pa)} &  \multicolumn{2}{c}{$\chi$ (-)} & \multicolumn{2}{c}{$D$ ($\times 10^{-7}$ m$^2$/s)} \\
		& \scriptsize $R=0.5\,\mu\rm{m}$ & \scriptsize $R=3.5\,\mu\rm{m}$ & \scriptsize $R=0.5\,\mu\rm{m}$ & \scriptsize $R=3.5\,\mu\rm{m}$ & \scriptsize $R=0.5\,\mu\rm{m}$ & \scriptsize $R=3.5\,\mu\rm{m}$ & \scriptsize $R=0.5\,\mu\rm{m}$ & \scriptsize $R=3.5\,\mu\rm{m}$ \\
		\midrule 
		1 & 1856 & 540 & 1125 & 357 & 0.42 & 0.48 & 0.02 & 0.30 \\ 
		2 & 804 & 192 & 1476 & 473 & 0.44 & 0.48 & 0.03 & 1.50 \\ 
		3 & 1045 & 305 & 1395 & 435 & 0.42 & 0.47 & 0.02 & 0.18 \\ 
		4 & 558 & 229 & 1558 & 461 & 0.45 & 0.48 & 0.09 & 1.18 \\ 
		5 & 862 & - & 1456 &  - & 0.42 & - & 0.01 & - \\ 
		\midrule 
		Mean & 1025 & 316 & 1402 & 431 & 0.43 & 0.48 & 0.03 & 0.79 \\ 
		\midrule 
		Std & 496 & 156 & 165 &  52 & 0.01 & 0.01 & 0.03 & 0.65 \\  
		\bottomrule
	\end{tabularx}
	\vspace*{8pt}
	\caption {Values of the shear modulus $G$, bulk modulus $K$, solvent-polymer matrix mixing parameter $\chi$ and water diffusivity $D$ of the basement membranes, as obtained from the numerical fitting procedure for two values of the indenter (radius $R=0.5$~\si{\micro\meter} and $R=3.5$~\si{\micro\meter}), and force setpoints ($F = 0.5$~nN and $F = 1$~nN), respectively. Partial data is available for BM5, due to the limited thickness of the membrane, which prevented indentation at the highest force setpoint.}
	\label{tab:values}
\end{table}

To rationalize the observed force relaxation and extract the material properties of the membranes, finite element simulations of the indentation tests were performed using a poroelastic model for the basement membrane as described in the Methods section. The computational domain depicted in Fig.~\ref{fig:simulations}b was set using information on the thickness from AFM-imaging for each membrane analyzed. As reported in Fig.~\ref{fig:simulations}c, indentation induces a localized solvent overpressure within the membrane of the order of $1.5$~$\mbox{kPa}$, which  drives fluid away from the indented region. The following solvent leakage from the free surface of the membrane relieves this overpressure and causes a reduction in the indentation force, until a new chemo-mechanical equilibrium is attained at the force plateau. This behavior is typically found in indentation tests of poroelastic materials \cite{Oyen2006,Hu2010}. 
Upon fitting the material parameters for each membrane to the datasets corresponding to $F = 1$~nN, $R = 3.5$~\si{\micro\meter} and $F = 0.5$~nN, $R = 0.5$~\si{\micro\meter}, the model captures quantitatively the measured average force-relaxation curves (Fig.~\ref{fig:simulations}d). The poroelastic properties of the membranes resulting from the fitting procedure are summarized in Table~\ref{tab:values}. We remark that the values of the shear and bulk moduli refer to the free swelling equilibrium, \textit{i.e.}, the undeformed membrane, and vary, for instance, with the hydration level \cite{Lucantonio2012}.

To validate the model, the force-relaxation curve for $F = 0.5$~nN, $R = 3.5$~\si{\micro\meter} was simulated, for each membrane, using the parameter values fitted to reproduce the ($F = 1$~nN, $R = 3.5$~\si{\micro\meter}) curve. For the given indenter size, the model captures the poroelastic dynamics for both force setpoints, which correspond to different maximum indentation depths (Fig.~\ref{fig:simulations}d). Additional comparison between experiments and simulations is reported in the Supporting Materials (Fig.~S1). 

Finally, in order to assess the physiological relevance of poroelastic flow, we performed a simulation to estimate the solvent flux through the basement membrane under a compressive load along the thickness direction. The load was chosen to have a peak amplitude of 500~Pa and a periodicity of 1~Hz (see Fig.~S3 of the Supporting Materials). Such forces occur {\it e.g.} during physical exercise \cite{Lu} and even higher pressures have been recorded during milk ejection \cite{Elendorff,Kent}. The relative volume change experienced by the membrane in this case reached values of 22\%, indicating that the fluid exchange due to realistic loads is substantial and likely to influence transport of all biomolecules throughout the basement membrane.

\section*{Discussion}
\subsection*{Age-dependent changes and permeation}
Upon BM dehydration, a tenfold decrease in BM thickness was observed in all the membranes analyzed. This result is well in line with previously published findings for the internal limiting membrane \cite{bm_H2O} and can be probably explained by the high quantity of water-retaining heparan sulfate proteoglycans (HSPGs) physiologically present in BMs. Although we cannot exclude that other basement membranes might react slightly differently to dehydration due to their different protein content, we can safely assume that water constitutes a fundamental component of BM architecture. Hence, any model of the mechanical material properties of BMs should - at least implicitly - incorporate the role of water. Here, we modeled the acinar BM as a hyperelastic matrix immersed in fluid.

To assess the porous nature of the {\it in-vitro} grown basement membranes, we performed immunostainings of what are believed to be the most important breast BM components, namely collagen IV and laminin-3A32.
Despite the fact that the collagen IV network does not significantly alter its characteristic pore size with maturation, previous experiments had highlighted a clear variation in the permeability properties of BMs over time: fluorescently labeled 40~kDa dextran molecules diffuse freely through low-matured BMs, but their permeation through highly-developed BMs displays a retardation effect that allowed to estimate a pore size of at least 9~nm \cite{paper_Aljona}. These observations, consistent with the notion of membrane thickening over time, indicate that the fine protein meshwork filling the larger collagen IV pores must become denser during acinar development. The very high values of water diffusivity reported here (comparable to those of certain hydrogels \cite{Hu2011}) are also well in line with the high dextran permeability already described for low-matured membranes \cite{paper_Aljona}. Additionally, the observation of age-dependent variations in BM thickness further validate the use of MCF10A as a physiologically relevant model system given that, {\it in-vivo}, the mammary duct BM is known to undergo significant alterations during the process of mammary gland morphogenesis. The formation of the ductal tree in fact involves a continuous reshaping of basement membranes, which can be as thin as 104 nm at the sites of ductal branching initiation \cite{mina_mmp} but reach up to 1.4~\si{\micro\meter} in thickness \cite{mina} along the flanks of the terminal end buds.

Interestingly, previous work from some of the authors could show that the disruption of the collagen IV meshwork completely abolished the size-dependent molecule retardation effect of fully-developed BMs \cite{paper_Aljona}. This result is in full agreement with the notion emerged from confocal microscopy imaging, namely that collagen IV acts as the main structural component of the protein network, despite the estimated presence of over 50 different proteins in BMs \cite{Kalluri}.

In light of earlier work \cite{Coll_2}, our results on the collagen IV structure are surprising. Instead of forming a network with mesh sizes of about 50 nm as is well-documented from rapid-freeze, deep-etch replication experiments \cite{Coll_1, Coll_IV_lens_capsule}, the collagen network appears spatially heterogeneous, with denser areas that leave space to pores as large as one micrometer. Although pores of 800 nm would be in line with the model for collagen IV assembly based on 7S and NC1 bonds originally proposed by K{\"u}hn \cite{Kuehn}, such structures have not been described elsewhere. One explanation for this discrepancy could be that the absence of a myoepithelial layer in the 3D cell culture model causes the absence of specific interactions necessary for the full formation of a physiological network (myoepithelial cells have been shown to secrete collagen IV and greatly help the structural arrangement of breast lobules \cite{myoepith_cells}).
On the other hand, it is well established that BMs of tissue in the human body requiring disintegration or massive remodeling (such as amniotic BMs) are significantly less crosslinked and hence more susceptible to proteolysis than other BMs \cite{Kalluri}. Given the extensive remodeling that breast glands undergo during puberty, pregnancy and lactation, it is plausible that spatial collagen distribution and superstructure in the BM could be highly heterogeneous. Only comparison with super-resolution imaging of native breast BMs at different developmental stages could clarify this matter; however, such issues do not affect the validity of our poroelastic analysis, which is independent of the exact molecular architecture of the BM network.

In the future, it would be interesting to investigate the spatial localization of other prominent BM molecules as well, for instance the network linkers nidogen and entactin. 
In the present study, we focused on the non-network forming laminin 3A32 isoform due to its high relevance for breast gland BMs. Uniform secretion of laminin-3A32, in fact, is a marker of correct MCF10A morphogenesis \cite{rotation}, and its loss a classic hallmark of breast carcinoma \cite{lam5_breast_1, lam5_breast_2}. Laminin-3A32 is also a fundamental coordinator of BM network structure thanks to its capacity to bind collagens and nidogen, as well as the $\alpha$6$\beta$4 and $\alpha$3$\beta$1 integrins of breast gland epithelia \cite{Yuri}.

The here reported spatial colocalization of laminin and perlecan is in line with previous findings \cite{lam_perl_1}; trimolecular complexes of laminin, perlecan and dystroglycan, for instance, have already been described within cerebral cortex microvessels \cite{lam_perl_LARGE}.
In lung cells, a laminin-311-perlecan complex is necessary for mechanotransduction as it activates the MAPK pathway upon cyclic stretching \cite{lam_perl_lungs}. Given the high level of mechanical activity of epithelial breast gland cells, it will be crucial to explore the mechanosensing role of laminins also in this context.

\subsection*{Poroelasticity and structure of breast gland BM at different scales}
The AFM force relaxation data reported offer a very strong indication that breast gland BMs mechanically behave as a poroelastic material. The excellent match of the FE simulations with the experimental relaxation curves allowed for the first quantification not only of BM stiffness, but also of the diffusivity of water through the membranes. In essence, a poroelastic formulation allows to describe the coupling of BM transport properties with matrix elasticity; information that would just be lost upon simplistically modelling the membranes as purely elastic objects, like often found in literature \cite{bm4,bm_H2O,cornea}.
 
An important validation of the proposed mechanism came from comparing relaxation curves having different force peaks but measured with the same indenter. Parameters extracted by FE simulations and then used to predict the membrane relaxation at different initial forces described the experimental curves with excellent agreement. In particular, our results show that the relaxation curves for different indentation depths can be fitted with a poroelastic model using the same value for the diffusivity $D$. Therefore, for a given indenter size, the characteristic time scales as $a^2/D$, where $a$ is the contact radius depending on the indenter force (and hence on the indentation depth). A purely viscoelastic relaxation mechanism, on the other hand, would be independent of length scales \cite{Hu2011,Kalcioglu2012} and could explain the slight underestimation of the force peak present in some cases (see Fig.~S1), although this may also be due to tissue anisotropy \cite{Hatami-Marbini2016}. In conclusion, while we cannot exclude the additional presence of viscoelastic processes, the excellent agreement between numerical and experimental data supports the assumption that poroelasticity is the fundamental physical mechanism underpinning the force relaxation and the mechanical behavior of basement membranes.

A point of interest lies in the fact that the values of $G$ and $K$ obtained for the two indenter radii are shifted of a constant factor of about 3. Keeping in mind that part of this difference is certainly due to experimental uncertainty (each AFM cantilever, for instance, is subject to a certain calibration error), in general such a trend should not surprise. The mechanical properties of microtissue are known to vary according to the scale analyzed: upon indenting cartilage samples with microspherical indenters, for instance, a 100-fold variation in their dynamic elastic modulus has been observed \cite{cartilage} as compared to that obtained using sharp pyramidal tips. Similar studies are reported for bone \cite{bone} and liver \cite{liver} samples, but the effect is expected to be ubiquitous for all types of tissue. Here, given that the smaller indenter radius ($R=0.5$~\si{\micro\meter}) is exactly comparable to the characteristic length of the collagen IV meshwork, we can understand why curves recorded with this cantilever gave higher stiffness values: when indenting with the 3.5~\si{\micro\meter} bead, local heterogeneities at the nanoscale level are not resolved, and the chance of indenting exclusively on thick fiber bundles decreases. 

Differences in the permeability of bone and cartilage tissue as assessed with nano- and micro-indentation tests have also been reported \cite{Galli} and are typically due to the different microstructure of tissue at different scales. Here, the variation of the diffusivity with indenter size may be explained by the associated change in solvent pressure acting on the polymer matrix. Specifically, diffusivity should decrease with solvent pressure because of the collapse of collagen pores under compression \cite{Chandran2004}. Consistent with this interpretation, we have found a sharp decrease in the fitted values of $D$ as the maximum solvent pressure attained during each indentation test increases (see Fig.~S2). Indeed, in collagen-based tissues, an exponential dependence of the permeability $k$ (which is proportional to the diffusivity $D$) on the volumetric strain is usually assumed \citep{Ateshian1997}. This implies that $k$ strongly decreases with volume shrinkage as a consequence of the pressure acting on the matrix. From the relation $D=\eta k \mc{R}T/(\Omega \phi_{\rm o})$ \cite{Yasuda1971}, where $\eta$ is the viscosity of water and $\phi_{\rm o}$ is its volume fraction in the membrane at equilibrium, we can estimate the average permeability to be $k \approx 6\times10^{-19}$~$\mbox{m}^2$. This value corresponds to a pore size $d \approx 4$~nm, assuming the relation $k = d^2/32$ that holds for laminar flow through a cylindrical tube of diameter $d$ \cite{LandauLifshitz}. This estimate for $d$ is in the same order of magnitude as values estimated in previous work \cite{paper_Aljona}.

Permeation in the breast is a highly physiological function. Even though basement membranes act as crucial gatekeepers for molecule diffusion and permeation, though, their porosity has so far received relatively little attention. A notable exception is given by kidney glomerular BMs, whose filtration properties have been studied for decades \cite{GBM}: their permeability barrier lies in the range of 40-200 kDa \cite{plos_10}, and their protein architecture is characterized down to the nanoscale-level \cite{GBM_nano}.
In the breast, the frequency and extent of milk removal from the lactating mammary gland determines the rate of further milk secretion thanks to the regulatory action of the FIL protein (Feedback Inhibitor of Lactation), a 7.6 kDa whey polypeptide \cite{milk_FIL}. Other hormones and signaling factors also need to normally cross the BM in order to perform their biological function during the different developmental stages, and even maternal dietary proteins of up to 43 kDa have been shown to permeate to the alveolar units of mammary glands in order to be passed to the infant via milk suckling \cite{proteins_passage_milk}. Hence, a better insight in the filtration properties of the BM would have significant implications for drug delivery. In previous experiments, we identified a permeation barrier for MCF10A BMs in the range of 40 kDa \cite{paper_Aljona}.
Now we interestingly observe that a polymer network behaving as a poroelastic material would also contribute to the breast gland flow properties by increasing the release of trapped fluid upon increased external pressure. 

To estimate this effect, we performed a simulation in which a compressive load of 500 Pa with a periodicity of 1~Hz was applied to the BM along the thickness direction; such values have been associated with the rhythmic dynamic pressures experienced by breast glands during activities such as running \cite{Lu} and become even higher during suckling and lactation \cite{Elendorff,Kent}. The corresponding relative volume variation experienced by the membrane reached a peak value of 22\% (see Fig.~S3). Given that intramammary pressures recorded {\it in-vivo} during suckling are reported to be in the range of 2-6 kPa \cite{Elendorff, Kent}, the effective flow and hence transport of molecules in humans might be even higher than that estimated in our simulation. Even though calculations of specific proteins' fluxes are beyond the scope of this work, such substantial fluid exchange is most likely to influence the transport of all biomolecules throughout the basement membrane.

The delivery of hormones such as prolactin to the milk-secreting inner layer of epithelial cells, therefore, is likely not only guided by free diffusion, but could also be tuned in response to the varying tension and strain within tissue, which are especially high in the lactating breast. 
A confirmation of this mechanism comes from the theoretical understanding that also for other poroelastic materials, such as polymer gels, the stress-diffusion coupling indeed affects the kinetics of solvent diffusion within the polymer network \cite{Lucantonio2014}.

\section*{Conclusion}
In this study, we analyzed the nanoscale topography of MCF-10A-derived breast gland basement membranes and assessed their mechanical properties by means of AFM nanoindentation. We found that the mechanical response of isolated BMs can be described in terms of a poroelastic material model, in line with the reported membrane's structural composition ({\it i.e.}, a porous protein meshwork immersed in fluid). Additionally, we report differences in the permeability and mechanical properties of BMs at different scales of tissue architecture (namely, at the nano- and micro-scale), as already well established for ECM-rich tissues such as cartilage \cite{Galli}.
Even at the model system level, our understanding of the complex interplay between ECM environment and breast gland regulation is still at its infancy, despite carrying profound implications for tissue development and cancer spreading. Serving both as mechanical scaffolds and as pressure-dependent reservoirs for various signaling molecules, BMs lie at a crucial interface linking the biochemical alterations known to accompany cancer development and progression with the biomechanical response of breast gland tissue. 
In the future, a more comprehensive study of BM properties variation with developmental stage and pathogenicity level would be of prime interest, although this would require the more complex analysis of tissue-isolated membranes. Future studies on the mechanics of BMs should also focus on their fracture properties and toughening mechanisms in relation to poroelasticity \cite{Lucantonio2015PRL,Noselli2016}, as these may also have an impact on cancer spreading.
Focusing on the initial analysis of {\it in-vitro} secreted model systems, our data present the first detailed characterization of the architecture of breast gland BMs and, based on their structure, the identification of the fundamental mechanism describing material response ({\it i.e.}, poroelasticity). Hopefully, this will offer a step forward towards a deeper understanding of breast gland tissue mechanoregulation.

\section*{Author Contributions}

G.F. performed research, analyzed data, wrote the paper. A.L. performed research, analyzed data, wrote the paper. N.H. performed research. E.N. performed research. B.H. performed research. A.D.S. designed research. R.M. designed research, wrote the paper.

\section*{Acknowledgments}

A.D.S. and R.M. thank the Isaac Newton Institute for Mathematical Sciences for its hospitality during the program
'Coupling geometric partial differential equations with physics for cell morphology, motility and pattern formation' (CGPW02) supported by EPSRC grant no.~EP/K032208/1. 
A.L. and A.D.S. acknowledge support from the European Research Council through AdG-340685 - MicroMotility. We thank G. Dreissen (ICS-7) for the collagen filaments masking algorithm, F. Kumpfe (JPK) for precious technical support, F. Santoro (IIT) for helpful discussions on SEM imaging protocols and the Helmholtz Nano Facility (HNF) \cite{HNF} for access and expert assistance.

\section*{Supporting citations}
References \cite{doi_gel_2009, comsol} appear in the Supporting Material.

\end{document}